# Gate-tunable conducting oxide metasurfaces

Yao-Wei Huang[1,†], Ho Wai Howard Lee[1,2,†,+], Ruzan Sokhoyan[1], Ragip Pala[1,2], Krishnan Thyagarajan[1,2], Seunghoon Han[1,3], Din Ping Tsai[4], and Harry A. Atwater[1,2,*]

[1]Thomas J. Watson Laboratories of Applied Physics, California Institute of Technology, Pasadena, California 91125, United States
[2]Kavli Nanoscience Institute, California Institute of Technology, Pasadena, California 91125, United States
[3]Samsung Advanced Institute of Technology, Samsung Electronics, Suwon-si, Gyeonggi-do 443-803, Republic of Korea
[4]Research Center for Applied Sciences, Academia Sinica, Taipei 11529, Taiwan

## Abstract
Metasurfaces composed of planar arrays of sub-wavelength artificial structures show promise for extraordinary light manipulation; they have yielded novel ultrathin optical components such as flat lenses, wave plates, holographic surfaces and orbital angular momentum manipulation and detection over a broad range of electromagnetic spectrum. However the optical properties of metasurfaces developed to date do not allow for versatile tunability of reflected or transmitted wave amplitude and phase after fabrication, thus limiting their use in a wide range of applications. Here, we experimentally demonstrate a gate-tunable metasurface that enables dynamic electrical control of the ***phase*** and ***amplitude*** of the plane wave reflected from the metasurface. Tunability arises from field-effect modulation of the complex refractive index of conducting oxide layers incorporated into metasurface antenna elements which are configured in a reflectarray geometry. We measure a phase shift of $\pi$ and ~ 30% change in the reflectance by applying 2.5 V gate bias. Additionally, we demonstrate modulation at frequencies exceeding 10 MHz, and electrical switching of +/-1 order diffracted beams by electrical control over subgroups of metasurface elements, a basic requirement for electrically tunable beam-steering phased array metasurfaces. The proposed tunable metasurface design with high optical quality and high speed dynamic phase modulation suggests applications in next generation ultrathin optical components for imaging and sensing technologies, such as reconfigurable beam steering devices, dynamic holograms, tunable ultrathin lens, nano-projectors, and nanoscale spatial light modulators. Importantly, our design allows complete integration with electronics and hence electrical addressability of individual metasurface elements.

## Introduction
Metasurfaces are arrays of subwavelength elements in which each element is configured to control the phase and amplitude of the transmitted, reflected, and scattered light [1-3]. By controlling the phase shift and amplitude change imposed by each metasurface element, phased arrays are in principle achievable that would enable complex wavefront engineering. In the last four years, significant advances have occurred, resulting in a generalized form of Snell's law with ordinary and anomalous refraction, metasurfaces that refract and focus light, enabling applications such as holograms [4-7], optical vortex generation/detection, ultrathin focusing lens, photonic spin Hall effect [8-11], among others. Metasurfaces can also control beam polarization, yielding linearly, elliptically, circularly or radially polarized light [4, 12]. Due to the flat nature of metasurfaces (typical thickness < 100nm), conventional three-dimensional optical elements such as prisms or lenses can be replaced by their flat, low-profile analogs. An important practical consequence of the flat geometry is that metasurfaces are straightforwardly fabricated via planar lithographic processing.

Passive metasurfaces have been demonstrated in the spectral region ranging from visible to microwave frequencies. An as-yet unrealized milestone in the field is to achieve an actively tunable metasurface with arbitrary control of phase and amplitude of individual antenna elements, by post-fabrication electrical modulation which would enable dynamical wavefront control in thin flat optical devices, such as dynamical beam steering, reconfigurable imaging, tunable ultrathin lens, and high capacity data storage. There have been number of attempts to actively control the overall response of the metasurface by using various physical mechanisms [13-22]. It has been shown that one can actively control transmittance or reflectance of the impinging light. However to date there has been no comprehensive experimental validation of the promise for dynamical control of both the wave *phase* and the *amplitude* either in reflection or transmission from metasurfaces operating in the visible or near-IR. Moreover, it is worth mentioning that previous works on tunable metasurfaces show active control of amplitude over overall response of the metasurface without addressing the issue of individual control of metasurface elements. Here, we experimentally demonstrate independent electrical addressability to subgroups of three metasurface elements. This results in electrical switching of +/-1 order diffracted beams. Our design, in principle, allows addressing individual metasurface elements and complete integration with electronics.

Among various physical mechanisms for modulation of complex refractive index, field-effect modulation is a distinctly attractive approach because of the combined advantages of high speed modulation of large numbers of individual metasurface elements and extremely low power dissipation. Field-effect modulation is ubiquitous in semiconductor electronics, and is the principle underlying contemporary low power integrated circuit performance. Based on formation of charge depletion or accumulation regions in doped semiconductors (e.g., metal-oxide-semiconductor field-effect transistors and thin-film transistors) field effect modulation can provide a sufficiently large carrier density change in a heavily-doped semiconductor or conducting oxide, which results in a large variation of its complex refractive index in the charge accumulation or depletion regions [23, 24]. This phenomenon has been used to demonstrate electrically controlled plasmonic amplitude modulators in a metal-oxide-semiconductor (MOS) configuration, where coupling of modulator output power into a waveguide is controlled by an electrical bias across the metal-semiconductor field effect channel [23-25]. Transparent conducting oxide (TCO) materials [26, 27], such as indium tin oxide (ITO), have also been used as the active semiconductor layer [23, 25, 28-31]. The results reported by these authors suggest that applying an electrical bias between metal and ITO changes the sign of $\varepsilon_r$, the real part of the dielectric permittivity of ITO in the accumulation layer, from positive to negative. When $|\varepsilon_r|$ is in the epsilon-near-zero (ENZ) region, that is $-1<\varepsilon_r<1$, a large electric field enhancement occurs in the accumulation layer for near-infrared wavelengths [21, 25, 32, 33], providing an efficient way to electrically modulate the optical properties of nanophotonic devices with high modulation speed and low power consumption.

Here we integrate field-effect tunable materials with metasurfaces to demonstrate a dynamically tunable metasurface that allows for the first time active control of both reflected light *phase* and *amplitude* at near infrared wavelengths for applications such as dynamic beam steering. The metasurface we study consists of a gold back plane, an ITO layer followed by an aluminum oxide layer on which we pattern a gold nanoantenna array (Fig. 1a). The identical antennas are connected either to right or left external gold electrodes to create electrical gates. Unlike previous optical frequency metasurfaces which utilized variations in the antenna geometry or orientation to impose different phase shifts for each antenna element, the metasurface we study here is periodic, and a dynamic control over the phase shift by each metasurface element is achieved by applying bias voltages to combinations of adjacent antenna electrodes. Each metasurface antenna element is effectively an MOS capacitor with the Au antenna serving as a gate and ITO functioning as a field effect channel (Inset of Fig. 1a). When applying an electrical bias between the antenna gate and the underlying ground plane, the carrier concentration at the $Al_2O_3$/ITO interface increases or decreases by forming a charge accumulation or depletion layer. This results in modulation of the complex permittivity of ITO, which alters the interaction of the incident light with the antenna and modulates the reflection from the surface. The variation of the reflected phase and amplitude at each antenna element is amplified when the real part of the dielectric permittivity in the ITO accumulation layer changes its sign from positive to negative (Fig 1B). Our field effect electrostatics calculations indicate that this condition is satisfied when the background carrier concentration in ITO is $N_0 = 3\times10^{20}$ cm$^{-3}$, with a 5-fold increase in the carrier concentration at the Al2O3/ITO interface under an applied bias increase from 0 to 6 V.

Figure 1b (bottom) shows the calculated $\varepsilon_r$ for different applied biases as a function of distance from the $Al_2O_3$/ITO interface at an operation wavelength of 1500 nm. The dielectric permittivity of ITO substantially changes over the region within 1.5 nm of the $Al_2O_3$/ITO interface due to the formation of accumulation layer. The gray area highlights the ENZ region of ITO where $\varepsilon_r$ acquires values between 1 and -1. Under a positive bias, the value of $\varepsilon_r$ at the $Al_2O_3$/ITO interface decreases, reaching the ENZ condition at an applied bias of 2.5 V. The thickness of the ENZ region can be estimated as 0.7 nm at applied bias of 6 V. Importantly, when the ENZ condition holds, a large electric field enhancement is generated in the accumulation layer of ITO. This can be readily understood from the boundary condition imposing continuity of the normal component of electric displacement at the field effect dielectric/channel interface [25, 32, 33].

A photographic image of the final device, fabricated by multilayer deposition and e-beam lithography is shown in Figure 1c. One can visually distinguish the Au back plane, ITO, electrical connections, and pads (for further fabrication details, see Supplementary section S1). Scanning electron microscope images of a stripe antenna structure are depicted in Figs. 1D and E. Adjacent stripe antennas are connected electrically in groups of three so that each group is connected to a different external gold pad. The electrical pads are wire connected to a compact chip carrier and circuit board for electrical gating.

### Coupling of ENZ resonance via conducting oxide field-effect.
Using finite element electromagnetic simulation methods, we simulated reflectance and phase modulation of the periodically patterned antenna structure under normal incidence illumination with a transverse magnetic (TM) polarization (H-field along the stripes). In our simulations we considered an array Au stripes of width, $w = 230$ nm, thickness, $t = 50$ nm, and periodicity of $p = 400$ nm. The stripe array is on a 5 nm $Al_2O_3$ layer placed above a 17 nm ITO layer and a 100 nm Au back plane (Fig. 1a). Figures 2a and 2b show the reflection and phase shift spectra as a function of applied voltage. The black dashed lines indicate the ENZ region in the accumulation layer of ITO at the $Al_2O_3$/ITO interface, and the green line marks the position of a reflectance dip corresponding to the magnetic dipole plasmon resonance. With increasing gate bias, the magnetic dipole plasmon resonance couples to the ENZ region in the ITO accumulation layer which shifts the resonance and induces a significant phase shift in reflection.

The calculated phase as a function of wavelength under an applied bias from 0 V to 5.5 V is depicted in Fig. 2c. As one can see, there are two different regimes describing coupling of the plasmonic resonance with the ENZ region in ITO. For an increasing bias from 0 to 2.5V, the plasmonic resonance shifts to shorter wavelengths due to an increase in the carrier concentration and reduction of $\varepsilon_r$. As the resonance blue shifts, the phase shift at 1500 nm increases. In the case of applied bias of 3.5 V, $\varepsilon_r$ becomes zero, and applying a higher voltage, the plasmonic resonance shifts to longer wavelengths. This change in the sign of the resonant wavelength shift can be intuitively understood by looking at the behavior of the optical properties of the ITO layer. With increased bias, the carrier concentration in the ITO layer increases and starts functioning optically as a metal, thus shrinking the thickness of dielectric spacer. Interestingly, for applied biases exceeding 3.5 V, phase of the reflected beam decrease with increasing wavelength as opposed to the case of smaller applied bias (0 – 2.5 V), which shows an increase in the reflection phase with increasing with wavelength. Therefore there is a resulting phase shift larger than 180 degrees at an excitation wavelength of 1500 nm.

To gain further insight into the coupling of the magnetic dipole plasmon resonance with the ENZ region of the ITO, the distribution of electromagnetic field at different applied voltages is simulated (Fig. 2d, e). At 0 V, ITO optically behaves as a dielectric, thus the resonance shows a distribution of anti-parallel electric field and forms magnetic dipole between Au antenna and Au back plane. For the applied bias larger than 2.5 V, when the ENZ condition holds in the accumulation layer of the ITO, the enhancement of the z-component of the electric field ($E_z$) in the accumulation layer is observed (see second image in Fig. 2e) due to continuity of the normal component of electric displacement ($\varepsilon_\perp E_\perp$) at the interface of the two media. In case of applied voltage of 3.5 V where $1 > \varepsilon_r > 0$ in the accumulation layer, $E_z$ is enhanced in a direction parallel to the field in the $Al_2O_3$ and bulk ITO (second image in Fig. 2e). In contrast, for applied voltage of 5.5 V (third image in Fig. 2e), since the permittivity of ITO in some region of the accumulation layer is negative (metal-like), the $E_z$ component is antiparallel to the field in the $Al_2O_3$ and bulk ITO layers. The enhanced parallel and antiparallel $E_z$ due to coupling to ENZ region further modifies the strength of magnetic dipole and phase shift profile as can be seen in Fig. 2e, resulting in the large phase modulation via electrical gating.

### Results of tunable conducting oxide metasurfaces

The reflectance spectra and the metasurface phase shift are measured for different applied biases as shown in Fig. 3 (c.f., Supplementary section S2 for measurement setup). In Fig. 3a, we observe that the resonant reflectance minimum shifts to shorter wavelengths when increasing voltage as expected from the simulation results (Fig. 2a). The reflectance change (normalized to the reflectance without applied voltage) is as high as 28.9% for operating wavelengths near the resonant reflectance minimum under an applied bias of 2.5 V (Fig. 3b), indicating gate-actuated metasurface reflectance amplitude modulation. To measure the phase shift, a Michelson interferometer was used to observe interferometric fringes where the incident beam is positioned at the edge of the metasurface so that half of the beam is reflected from the metasurface and the other half is reflected from the bottom gold back plane, which acts as a built-in phase reference (Supplementary section S3-4). The phase shift measurement laser wavelength was $\lambda$ =1573 nm. Phase shifts were retrieved at different applied voltages by fitting and analyzing the interference fringes (Supplementary). The results are shown in Fig. 3c. It is clear that the phase shift increases with applied bias. A phase change of ~184 degrees is observed at applied bias of 2.5 V, which is in a good agreement with simulation results (inset of Fig. 3c). The slight discrepancy may be attributed to the small geometry, dielectric constant, and the intrinsic material properties (e.g., work function and effective mass of materials) difference between the simulated and experimental structures. We also applied negative bias voltages which, in contrast to the positive bias voltages, further deplete the ITO at the ITO/Al$_2$O$_3$ interface. In case of a negative applied bias, a red-shift of the resonant reflectance minimum is observed, indicating that we can both increase and decrease $\varepsilon_r$ near the ITO/Al$_2$O$_3$ interface by applying positive or negative biases, respectively. This excludes alternative interpretations of the observed resonance shift mechanisms including but not limited to thermal (Joule) heating (Supplementary section S8). We note that the 5 nm Al2O3 layer exhibits electrical breakdown at ~2.5−3 V.

To characterize the response frequency of our tunable metasurfaces, a 2 V bias with 0.5 - 10 MHz frequency was applied to the sample and a high speed InGaAs detector was used to detect the high frequency metasurface reflectance. To ensure high enough reflected light intensity, the width of the antenna was carefully designed such that the resonant reflectance minimum is located at ~ 1650 nm so we can record a sufficiently high light intensity at the telecommunication wavelengths (Supplementary section S9). As shown in the inset of Fig. 3d, high speed reflectance modulation is observed by applying 2 V of 500 kHz AC signal at the wavelength of 1515 nm (blue curve). Modulation speeds as high as 10 MHz were demonstrated (see Fig. 3d). Note that the waveform is distorted at the highest frequencies due to the detector bandwidth. The metasurface at 2 MHz frequency for different applied biases was also investigated and a ~ 15% of reflectance amplitude modulation is obtained with a 2 V applied voltage (Supplementary section S9).

The maximum speed of the modulation can be estimated using a simple device physics calculation, which gives a capacitance value of 14 fF/μm$^2$ per unit area (see Supplementary section S5). For our antenna array, this would yield capacitance and resistance values of 140 fF and 100 Ω per wire (area = 50 μm x 0.2 μm) respectively. Such a small capacitance enables modulation speeds up to 11 GHz and switching energies as low as 0.7 pJ/bit (this estimate does not include other sources of signal delay resulting from the wiring and RF probe connections). Note that the individual antenna can be implemented in a 2-dimensional antenna array with a smaller footprint (0.2 μm x 0.2 μm) and capacitance (0.5 fF). With 100 Ω resistance per pixel would enable 2.5 fJ switching energies per bit, and up to 3 THz modulation speeds (although the speed could be limited by interconnect with current speed up to ~ 100 GHz)[34]. We also anticipate that the modulation characteristics of the metasurfaces could be further improved by using alternative high-k gate dielectrics (Supplementary section S13).

As demonstrated in both experimental and simulated results, a large phase change (> 180 degrees) can be achieved. We further employ the tunable phase shift to develop an electrically driven dynamic phase grating. The voltage-dependent far-field intensity profiles of a metasurface phase grating with 64 unit cells consisting of identical patch antennas on Al$_2$O$_3$/ITO/Au planar layers were simulated. Fig. 4a shows the far field diffracted beam profile of a 2-level phase grating with a periodicity of Λ = 2.4 μm. At applied bias of 0 V, the diffracted beam shows a directional reflectance from the sample. While at the applied bias of 3.5 V, the 2-level phase grating creates two symmetric first-order diffracted beams with maximum intensity at angles of - 39 and 39 degrees due to the spatial symmetry of the structure. The experimentally measured far-field diffracted beam intensity is depicted in Fig. 4b. It can be seen that the ± first-order diffracted beams appear with applied voltage ~> 1.5 V while the zero-order diffracted beam intensity reduces with increasing voltage that agrees with the simulation results (Fig. 4a). These results manifest electrical tunability of reflection phase at a patch antenna level, which forms a fundamental basis for tunable optical phased array metasurfaces. Note that the slight discrepancy on the diffracted angles and beam widths between the simulated and measured results can be attributed to the non-parallel incident light excitation from the high NA objective and the slight non-uniformity of the sample. Importantly, the diffracted beam angle can also be varied by gating multiple antenna with different periodicities Λ for steering diffracted beam angles (Fig. 4 c-e). As shown in Fig. 4c, gating of 4, 3, 2 antennas periodically will lead to the grating periodicity of 3.2, 2.4, 1.6 μm, respectively (separation between each antenna is 400 nm), leading to the ability of steering diffracted beam angles (see schematics in Fig. 4c and d). To show the tunability of the diffracted angle for the first order diffraction, the far-field intensity of the first order reflected beam from the metasurface as a function of the incidence diffraction angles are simulated by gating the metasurface with 3.5 V (Fig. 4e). It is clear from the figure that the reflected beams are effectively steering to wide range of angles (> 40$^o$) by individually gating of 4, 3, 2 antennas to demonstrate the efficient nanoscale beam steering device. These efficient wide-angle electronically tunable beam steering components are necessary for the development of next generation ultrathin on-chip imaging and sensing devices, such as high resolution LIDAR devices and nanoscale spatial light modulators.

## Conclusions

In conclusion, we present a comprehensive experimental demonstration of a tunable metasurface in near infrared wavelength region. We control the phase and amplitude by gate-tunable conducting oxide field effect dynamical permittivity modulation. A phase shift of 184 degrees and reflectance change of ~ 30% were measured by applying 2.5 V gate bias. A modulation speed of up to 10 MHz (with potential modulation speed up to 11 GHz) and electrical beam-steering of +/-1 order diffracted beams were also demonstrated. In addition to the fundamental interest of tunable metasurfaces, these structures have many potential applications for future ultrathin optical components, such as dynamic holograms, tunable ultrathin lens, reconfigurable beam steering devices, nano-projectors, and nanoscale spatial light modulators.

## Acknowledgments
This work was supported by Samsung Electronics. The conducting oxide material synthesis design and characterization was supported by the U.S. Department of Energy (DOE) Office of Science grant DE-FG02-07ER46405 (K.T. and H.A.A.), and used facilities supported by the Kavli Nanoscience Institute (KNI) at Caltech. Y.W.H. and D.P.T. acknowledge the support from Ministry of Science and Technology, Taiwan (Grant numbers: 103-2911-I-002-594 and 104-2745-M-002-003-ASP). K. T. acknowledges funding from the Swiss National Science Foundation (Grant number: 151853).

## Author contributions
Y.H., H.W.L., R.S., and H.A.A. designed and conceived the experiments. Y.H. and H.W.L. fabricated the samples. Y.H. and H.W.L. developed the measurement setup. Y.H., H.W.L., and R.P. performed the experiments. Y.H., H.W.L., and K.T. performed materials characterizations. Y.H. and R.S. performed numerical simulations. Y.H., H.W.L., R.S., R.P., K.T., S.H., and H.A.A. wrote the paper. All authors discussed the results and commented on the manuscript.

† These authors contributed equally to this work.
* Corresponding author. E-mail: haa@caltech.edu
+ Current address: Department of Physics, Baylor University, Waco, Texas 76798, United States; Institute of Quantum Science and Engineering, Texas A&M, College Station, TX 77843-4242, United States

**Competing financial interests**
The authors declare no competing financial interests.

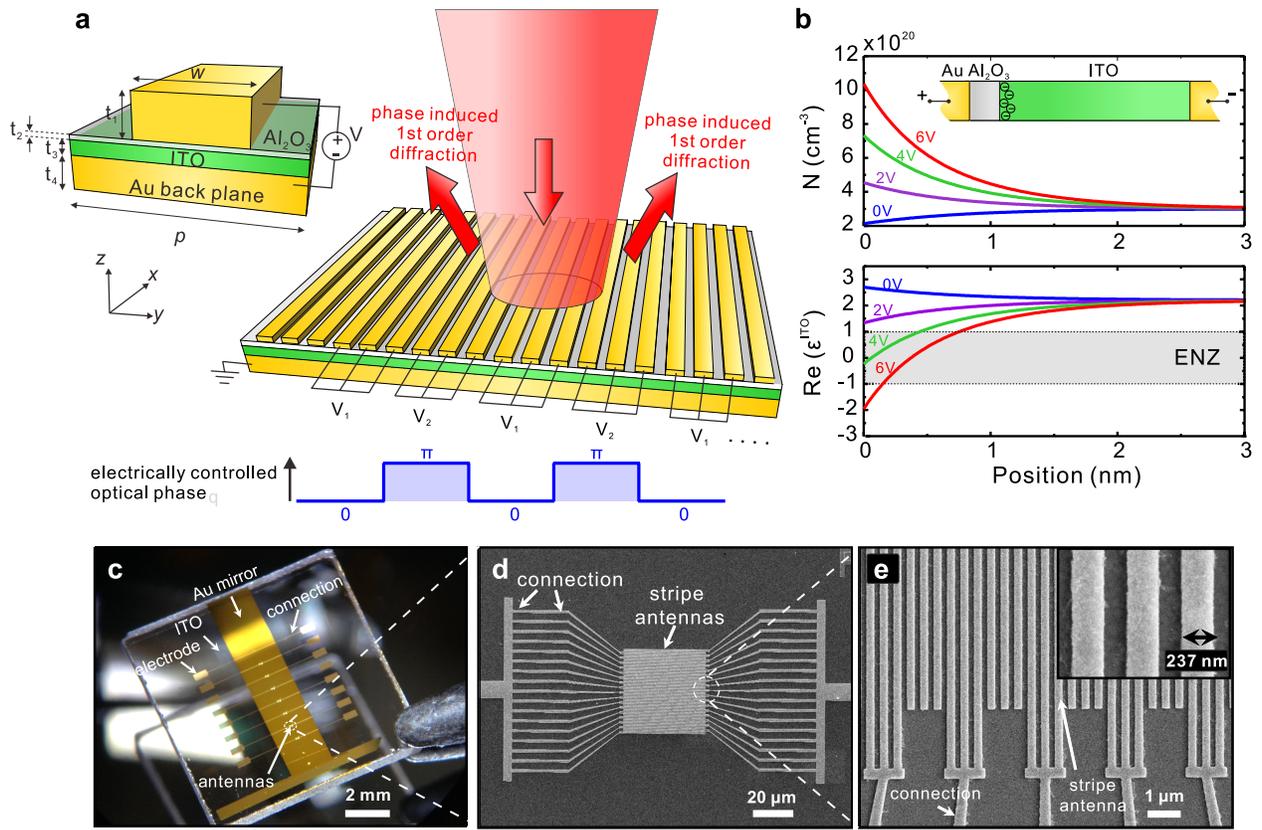

**Fig. 1. Gate-tunable metasurface.** (**a**) Schematic of the tunable metasurface. The structure consists of a quartz substrate, gold back plane, a thin ITO film, followed by a thin alumina film on which we pattern connected gold stripe nanoantenna array. Voltage is applied between the stripe antenna and the bottom gold resulting in the formation of charge accumulation at the $Al_2O_3$/ITO interface. Unit cell dimensions are chosen as follows: width of stripe antenna $w$ = 230 nm, thicknesses of the stripe antenna, $Al_2O_3$, ITO, and Au back plane - $t_1$ = 50 nm, $t_2$ = 5 nm, $t_3$ = 17 nm, $t_4$ = 100 nm, respectively. The periodicity of the unit cell is $p$ = 400 nm. (**b**) Operation principle of the tunable metasurface is based on MOS field-effect dynamics. When applied voltage is sufficiently high, an electron accumulation region forms in ITO at the $Al_2O_3$/ITO interface (inset). (Top) Spatial distribution of the carrier concentration $N$ for different applied voltages. (Bottom) Real part of the dielectric permittivity of ITO $\varepsilon_r$ at wavelength of 1500 nm as a function of distance from $Al_2O_3$/ITO interface for different applied voltages. Gray area highlights the spatial region where $\varepsilon_r$ acquires values between 1 and -1 representing the epsilon-near-zero (ENZ) region. (**c**) Photographic image of the fabricated structure. For gate biasing, stripe antennas are connected to electrical pads. Scanning electron microscopy images of (**d**) the stripe antenna and connections, and (**e**) the close up of stripe antenna.

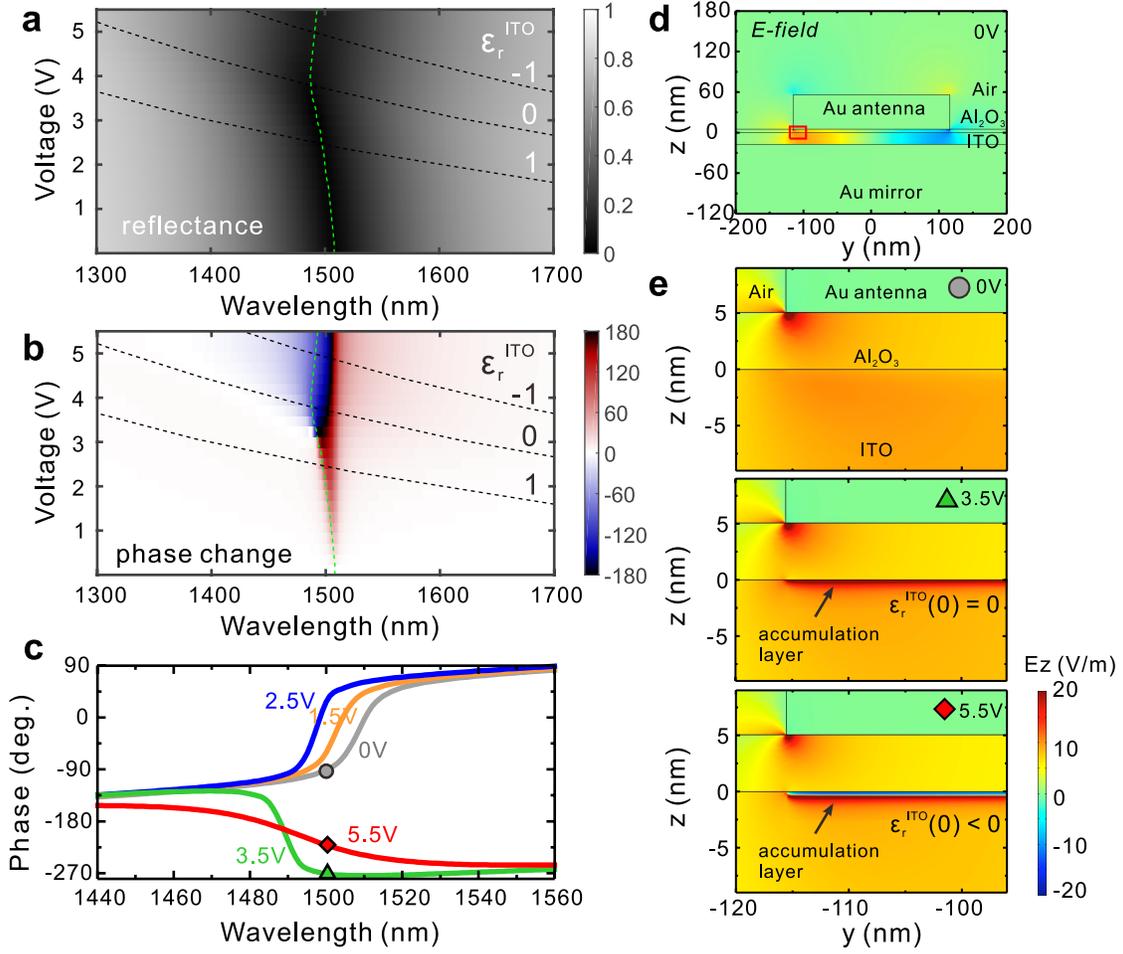

**Fig. 2. Gate-tunable metasurface based on interplay between magnetic plasmon resonance and ENZ region in ITO.** Simulated (**a**) reflectance and (**b**) phase shift due to gating as a function of wavelength and applied voltage. The phase shift is plotted with respect to the reflected phase from the metasurface without applied voltage. (**c**) Phase of the incoming plane wave acquired due to reflection from the metasurface as a function of wavelength for different values of applied voltage. (**d**) Spatial distribution of the $z$ component of the electric field $E_z$, and the magnified region near the ITO/Al$_2$O$_3$ interface for applied bias of 0 V, 3.5 V and 5.5 V at wavelength of $\lambda = 1500$ nm.

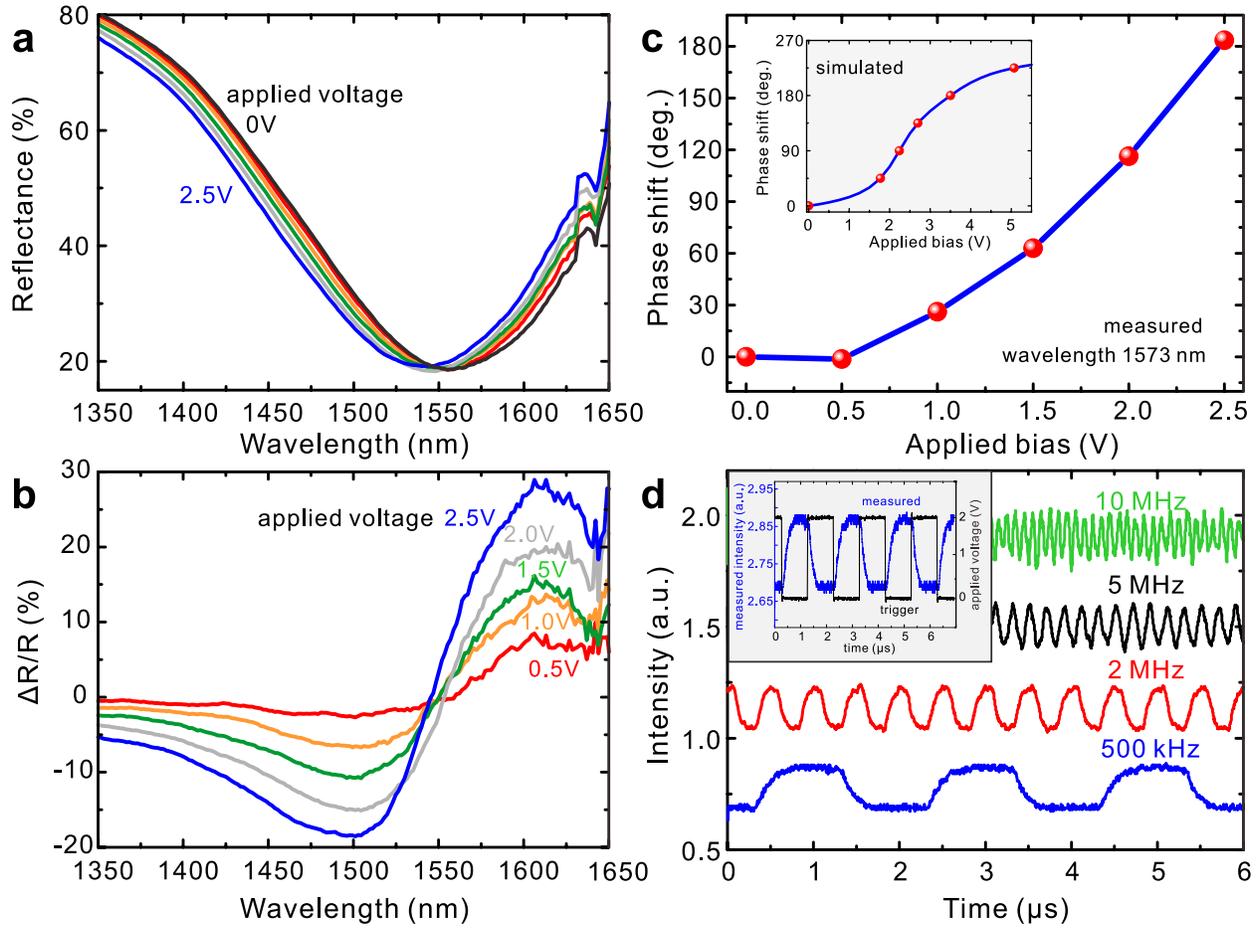

**Fig. 3. Experimental demonstration of π-phase modulation and amplitude modulation in tunable metasurfaces.** (**a**) Measured reflectance spectrum and (**b**) relative reflectance change from the metasurface for different applied voltages. (**c**) Experimentally derived phase shift as a function of applied bias for applied voltage between 0 V and 2.5 V. Inset: Simulated phase shift as a function of applied bias. (**d**) Measured high speed reflectance modulation with modulation frequencies from 500 kHz to 10 MHz. Inset: Measured AC modulated reflectance with operation speed of 500 kHz.

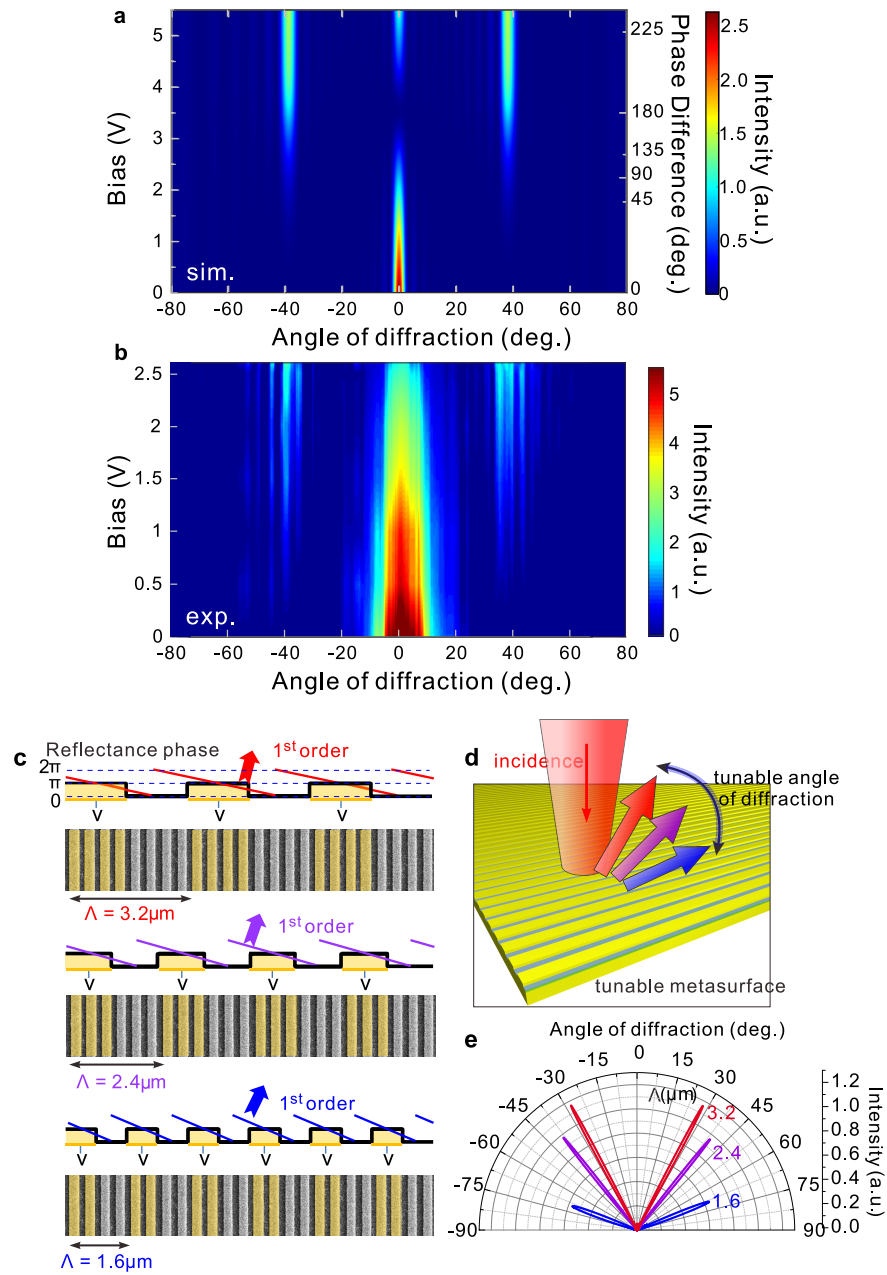

**Fig. 4. Dynamic phase grating with wide angular tunabilitiy. (a)** Simulated far-field intensity profiles of the diffracted beam versus applied bias. The color map shows far-field intensity of the light beam reflected from the metasurface as a function of diffraction angles and applied voltages for normal incidence. **(b)** Experimental measured far-field intensity profiles (detection angle up to ~ 50° with 0.75 NA objective lens). **(c), (d)** Schematic of steering diffracted beam angles via electrical gating of different number of antennas showing wide angular beam steering with high resolution. Periodically gating of 4, 3, 2 antennas effectively generate grating periodicities of 3.2, 2.4, 1.6 μm, respectively. **(e)** Simulated far-field intensity of the light beam reflected from the metasurface as a function of the diffraction angles for periodicities of 3.2, 2.4, 1.6 μm under a gate bias of 3.5 V. The device demonstrates an efficient beam steering with large range of angles (> 40°) with 2-4 periodically gated nano-antennas.